\RequirePackage{lineno} 
\documentclass[twocolumn,twoside,slac_two]{revtex4}
\usepackage{graphicx}
\usepackage{fancyhdr}
\usepackage{amsmath}
\newcommand{\bmu}{\boldsymbol{\mu}}
\newcommand{\bmhat}{\boldsymbol{\hat{\mu}}}
\newcommand{\btheta}{\boldsymbol{\theta}}
\newcommand{\bthat}{\boldsymbol{\hat{\theta}}}
\newcommand{\like}{\mathcal{L}}
\newcommand{\dset}{\boldsymbol{\mathcal{D}}}

\begin{document}
\title{Using Likelihood for Combined Data Set Analysis}

\author{B.~Anderson}
\affiliation{Wallenberg Academy Fellow.  Stockholm, Sweden.}
\author{J.~Chiang}
\affiliation{W. W. Hansen Experimental Physics Laboratory, Kavli Institute for Particle Astrophysics and Cosmology, Department of Physics and SLAC National Accelerator Laboratory, Stanford University, Stanford, CA 94305, USA}
\author{J.~Cohen-Tanugi}
\affiliation{Laboratoire Univers et Particules de Montpellier, Universit\'e Montpellier 2, CNRS/IN2P3, Montpellier, France}
\author{J.~Conrad}
\affiliation{Department of Physics, Stockholm University, AlbaNova, SE-106 91 Stockholm, Sweden}
\affiliation{The Oskar Klein Centre for Cosmoparticle Physics, AlbaNova, SE-106 91 Stockholm, Sweden}
\affiliation{Royal Swedish Academy of Sciences Research Fellow, funded by a grant from the K. A. Wallenberg Foundation}
\affiliation{The Royal Swedish Academy of Sciences, Box 50005, SE-104 05 Stockholm, Sweden}
\author{A.~Drlica-Wagner}
\affiliation{Center for Particle Astrophysics, Fermi National Accelerator Laboratory, Batavia, IL 60510, USA}
\author{M.~Llena~Garde}
\affiliation{Department of Physics, Stockholm University, AlbaNova, SE-106 91 Stockholm, Sweden}
\affiliation{The Oskar Klein Centre for Cosmoparticle Physics, AlbaNova, SE-106 91 Stockholm, Sweden}
\author{S.~Zimmer}
\affiliation{Department of Physics, Stockholm University, AlbaNova, SE-106 91 Stockholm, Sweden}
\affiliation{The Oskar Klein Centre for Cosmoparticle Physics, AlbaNova, SE-106 91 Stockholm, Sweden}
\author{on behalf of the Fermi LAT Collaboration}

\begin{abstract}
The joint likelihood is a simple extension of the standard likelihood formalism that enables the estimation of common parameters across disjoint datasets.  Joining the likelihood, rather than the data itself, means nuisance parameters can be dealt with independently.  Application of this technique, particularly to Fermi-LAT dwarf spheroidal analyses, has already been met with great success.  We present a description of the method's general implementation along with a toy Monte-Carlo study of its properties and limitations.
\end{abstract}

\maketitle

\thispagestyle{fancy}


\section{Introduction}

Several recent studies \citep{Ackermann:2011wa, 2012Sci...338.1190A, 2012JCAP...02..012F, Ackermann:2013yva, 2014ApJ...787...18A} by the LAT Collaboration successfully apply the joint likelihood technique, combining constraints for searches ranging from galaxy cluster emission to effects of large extra dimensions.  In the following, we introduce the technique from a more generic standpoint and compare/contrast it with other common methods of data combination.  We proceed with the aid of a toy Monte-Carlo (MC) to demonstrate the method's properties and explore some interesting behavior.

 \section{Likelihood analysis, Joint likelihood, and basic data stacking}
\subsection{Likelihood}\label{section:likelihood}
The likelihood incorporates information regarding both model and experiment into a function whose maximization provides an estimate of the true parameter values.  It can be expressed as 

\begin{equation}\label{eq:likelihood_def}
\like(\alpha|\mathcal{D})=P(\mathcal{D}|\alpha),
\end{equation}

\noindent
where $P$ is the probability of outcome, $\mathcal{D}$, given the parameter $\alpha$.  Parameters are often separated into those of interest, $\bmu$, and nuisance, $\btheta$, in order to profile or marginalize the latter.

We will focus on the specific form of $\cal L$ to be a binned Poisson probability function so that

\begin{equation}\label{eq:likelihood}
\like(\bmu,\btheta|\mathcal{D})=\prod_{k}\frac{\lambda_{k}^{n_{k}}e^{-\lambda_{k}}}{n_{k}!},
\end{equation}

\noindent
where the symbols $\lambda(\bmu,\btheta)$ and $n$ represent the predicted and observed counts in a given bin, $k$.  The parameters ($\bmhat,\bthat$) which yield the greatest value for ${\mathcal{L}}$ are known as the maximum likelihood estimate (MLE).  

When testing a hypothesis, the MLE likelihood must be compared with that of the null hypothesis, i.e. a model lacking the effect(s) of interest, where $\bmu\equiv \bmu_{0}$.  Typically, we compare the logarithms of the two likelihoods with a measure called the Test Statistic:

\begin{equation}
\rm TS = -2~\rm ln~\left(\frac{\like(\bmhat_{0},\bthat|\mathcal{D})}{\mathcal{L}(\bmhat,\bthat|\mathcal{D})}\right).
\end{equation}

\noindent
When its distribution is known, the TS can be mapped to a p-value associated with the alternative hypothesis.  In most cases, it obeys the asymptotic theorem \citep{Chernoff:1954} and follows a $\chi^{2}/2$ with degrees of freedom equal to the number of free signal parameters (assuming signal is constrained to be positive).  There are scenarios (some are mentioned in Section \ref{sec:caveats} and \cite{Ackermann:2013yva}) where this does not hold, and one must derive the TS behavior from a set of control data, e.g. with Monte-Carlo.  

\begin{figure}
\includegraphics[width=0.9\columnwidth]{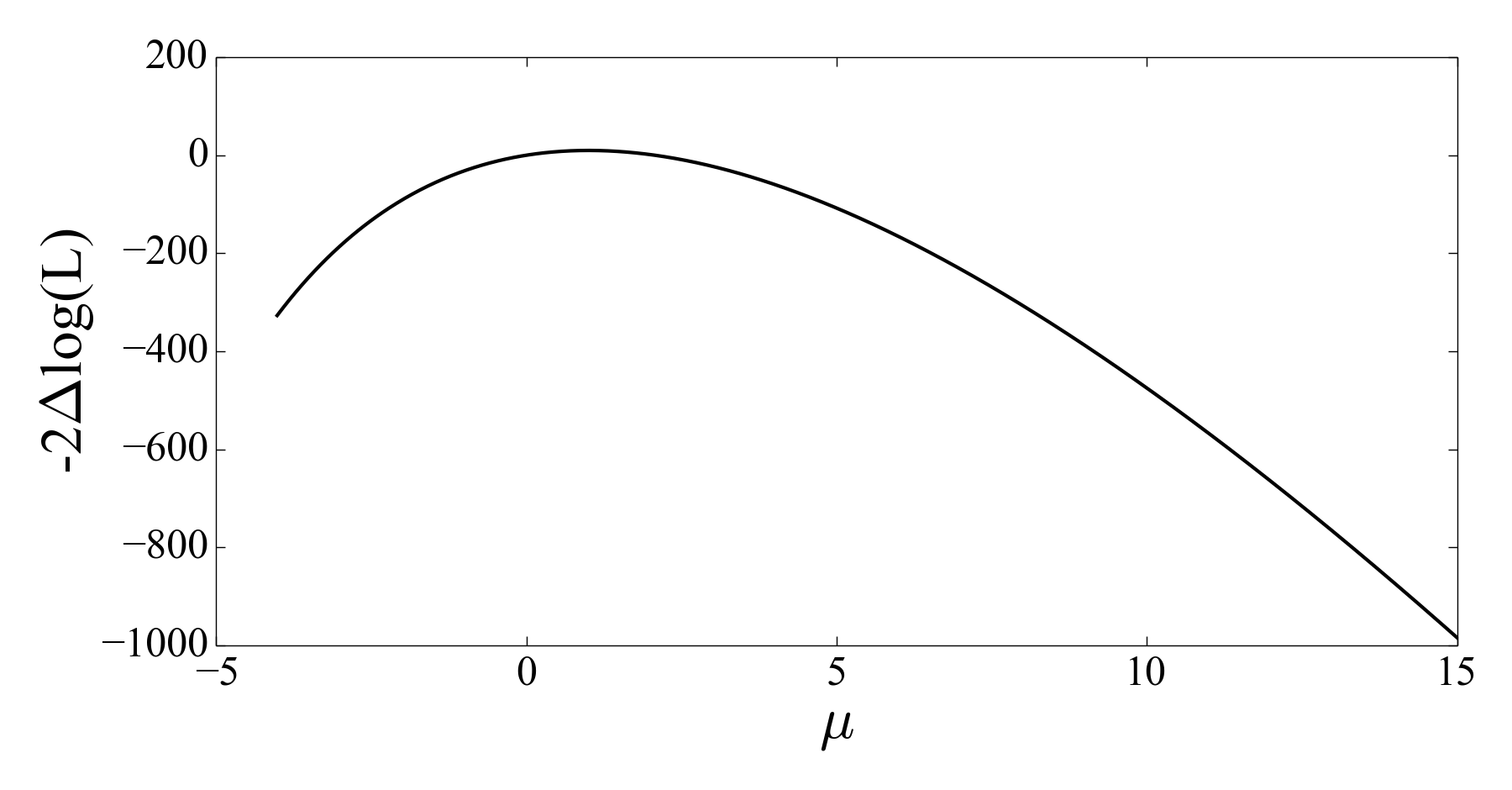}
\caption{A typical delta log-likelihood profile for a single-parameter source with no signal.\label{fig:profile}}
\end{figure}

Once the distributions are known, confidence intervals can be set for parameters by exploring the log-likelihood space surrounding the MLE.  Figure \ref{fig:profile} illustrates a typical delta log-likelihood profile for a simple system where there is only one free parameter that controls the strength of the new phenomenon.  Within the asymptotic regime, confidence limits would be set at levels corresponding to the $\chi^{2}$ probability density function, e.g. a difference of 2.71 from the maximum indicates 90\% one-sided coverage.

\subsection{Joint Likelihood}
To make use of a joint likelihood, one presumably has $N$ datasets which share some signal parameter(s), $\boldsymbol{\mu}$.  The procedure for joining is analogous to the way binned probabilities make up $\mathcal{L}$ --- simply take the product of each set's likelihood \citep{2015APh....62..165C}.  Explicitly,

\begin{equation}
\mathcal{L}_{\rm joint}(\bmu,\btheta | \dset) = \prod_{d=1}^{N}\mathcal{L}(\bmu_{d},\btheta_{d} | \mathcal{D}_{d})
\end{equation}

This construction is clean in the sense that the data sets remain disjoint.  Each could have different backgrounds, exposures, or have even come from different instruments.  All  these characteristics (nuisance parameters) are accounted for in the individual likelihoods.

\subsection{Data stacking}
Alternative methods for combining data exist, the most basic of which being to evaluate the likelihood of the data set union.  That is, instead of 

\begin{equation}
\prod_{d=1}^{N}\mathcal{L}(\bmu_{d},\btheta_{d} | \mathcal{D}_{d}),
\end{equation}

\noindent
we evaluate the stacked data likelihood:\footnote{A common example would be the addition of counts maps.}

\begin{equation}
\mathcal{L}(\bmu,\btheta | \cup \dset).
\end{equation}

Here, data sets are lumped together and then the hypothesis test is performed with respect to a model which is also the sum of individual expectations.  Switching to Pearson's $\chi^{2}$ and keeping the notation from the previous section, a stacked test statistic might look like this:

\begin{equation}
\chi^{2}_{\rm stack}(\bmu,\btheta)=\frac{\sum_{k}\left[\sum_{d}n_{d,k}-\sum_{d}\lambda_{d,k}(\bmu,\btheta)\right]^{2}}{\sum_{d,k}\lambda_{d,k}(\bmu,\btheta)}
\end{equation}

\noindent
As before, parameters are adjusted to optimize (in this case minimize) the $\chi^{2}$.  Significance and confidence intervals are directly interpreted according to the expected probability density function.  

Although easily done, it is not difficult to envision problems with such a strategy.  Data sets with weak signal-to-noise wash out when combined with those which are larger, though not necessarily more constraining.  This method throws away information and is therefore not optimal.  

One can do better by combining \emph{residuals}, i.e.

\begin{equation}\label{eq:chi2resid}
\chi^{2}_{\rm resid}(\bmu,\btheta)=\sum_{d,k}\frac{\left[n_{d,k}-\lambda_{d,k}(\bmu,\btheta)\right]^{2}}{\lambda_{d,k}(\bmu,\btheta)}
\end{equation}

\noindent
This is a much more viable alternative to the joint likelihood method.  Depending on the situation, however, its implementation can be tricky.  For example, suppose that the predicted number of events also depends on some nuisance parameter, $\epsilon$ (e.g. time or exposure).  Uncertainties on this parameter can be accounted for by adding an additional term to $\chi^2_{\rm resid}$ if they can be modeled as Gaussian. If not, there is no obvious way to include them in the data stacking approach, whereas modifying the likelihood is straightforward for any known model of the nuisance parameter uncertainty.

\section{Properties}
\subsection{Toy Model}
To illustrate the fundamental properties of the method, we employ a simple toy MC model for combining constraints: single-bin data sets with Poisson counts generated according to

\begin{eqnarray}\label{eq:expected}
\lambda_{d}(\bmu,\btheta)&=&\mu\cdot s_{d}+b_{d}\\
\bmu&=&\{ \mu,s_{d}\} \nonumber \\											\btheta&=&\{ b_{d}\} \nonumber
\end{eqnarray}

\noindent
Each set may have a different number of total events and has a background determined by the nuisance parameter, $b$.  Signal is determined from an individual, $s_{d}$, and common scale factor parameter, $\mu$.  The latter is the value we wish to estimate.  
 
\subsection{Confidence Interval, Coverage, and Power.}
\begin{figure}
\includegraphics[width=0.9\columnwidth]{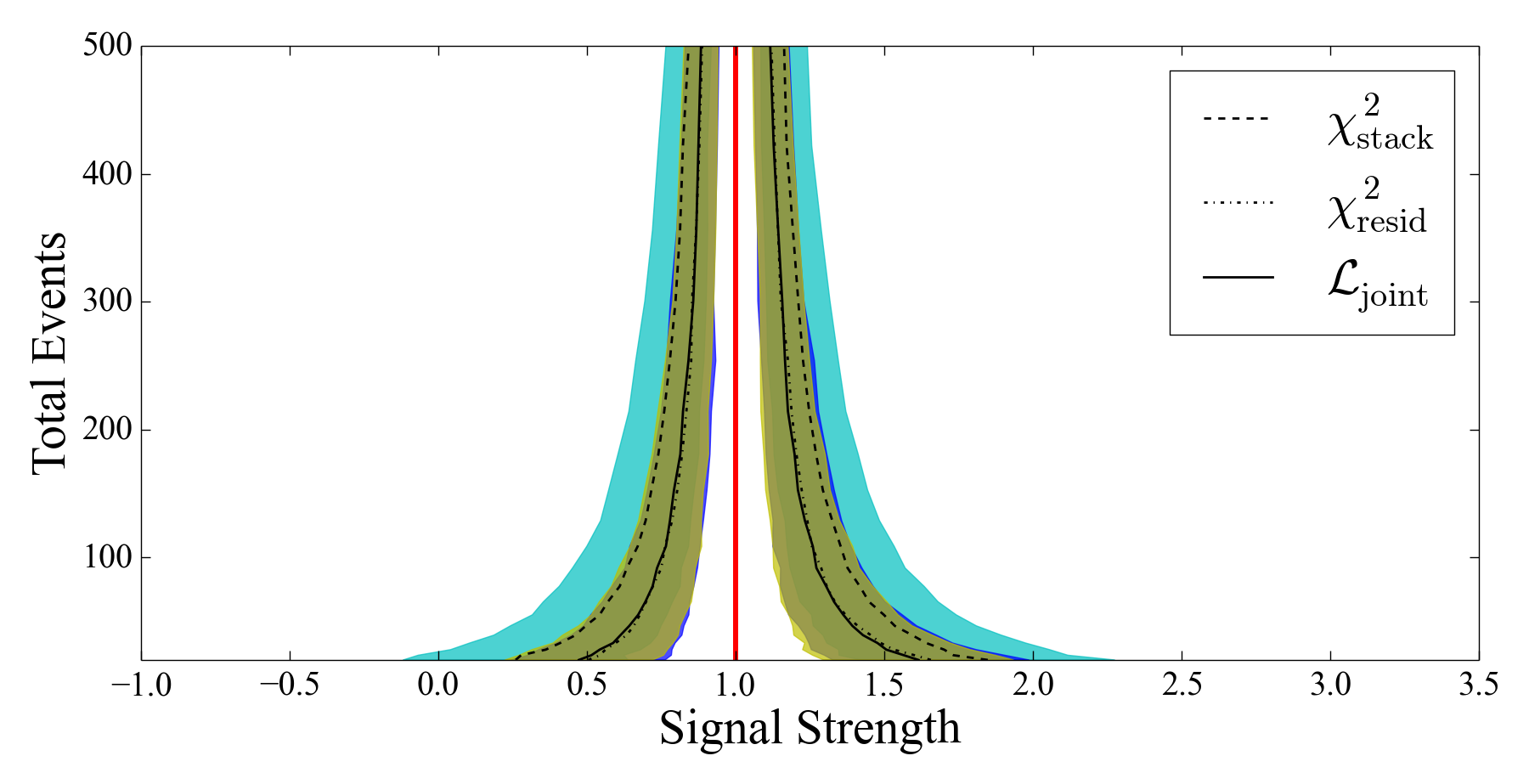}
\caption{Confidence intervals on the shared signal parameter, $\mu$, derived for two single-bin data sets with signal-to-background ratios of 1:1 and 1:10.  For an increasing number of total counts, 1000 MC realizations determine the median intervals for each method of combination.  Bands represent the 68\% containment among realizations.\label{fig:stack_limits}}
\end{figure}

As a starting point, we investigate the combination of two sets with signal-to-background ratios of 1:1 and 1:10.  Fig. \ref{fig:stack_limits} illustrates how the confidence intervals behave as a function of total events for the $\chi^{2}_{\rm stack}$, $\chi^{2}_{\rm resid}$, and $\mathcal{L}_{\rm joint}$ formulations.  In all scenarios, the coverage adheres to the nominal value and the limits improve in approximate proportion to the square root of the set size.  As expected, we see that the $\chi^{2}$ formed from residuals out-performs the simple stack (by yielding a tighter interval), and matches the joint likelihood.

\begin{figure}
\includegraphics[width=0.9\columnwidth]{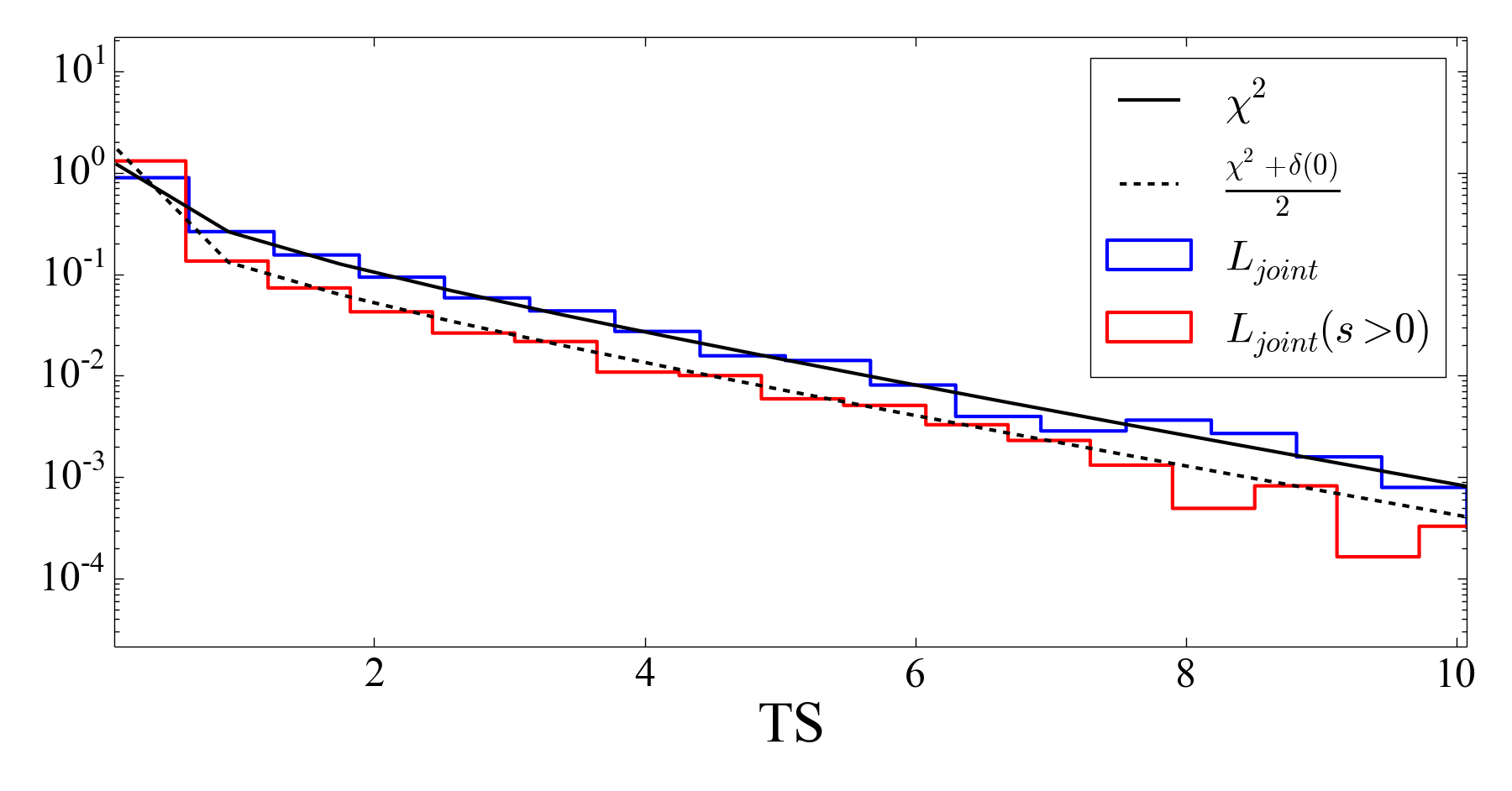}
\caption{TS distribution of two-set joint likelihood for both unconstrained and positive-only signal fits, along with the corresponding expected asymptotic distributions.\label{fig:null}}
\end{figure}

The TS distribution of the two-set joint likelihood is depicted in Fig. \ref{fig:null}.  Note that the distribution is halved (with the remaining stacked at zero TS) when the signal parameter is constrained to be greater than zero.  

\subsection{Effects of Additional Data Sets}
\begin{figure}
\includegraphics[width=0.9\columnwidth]{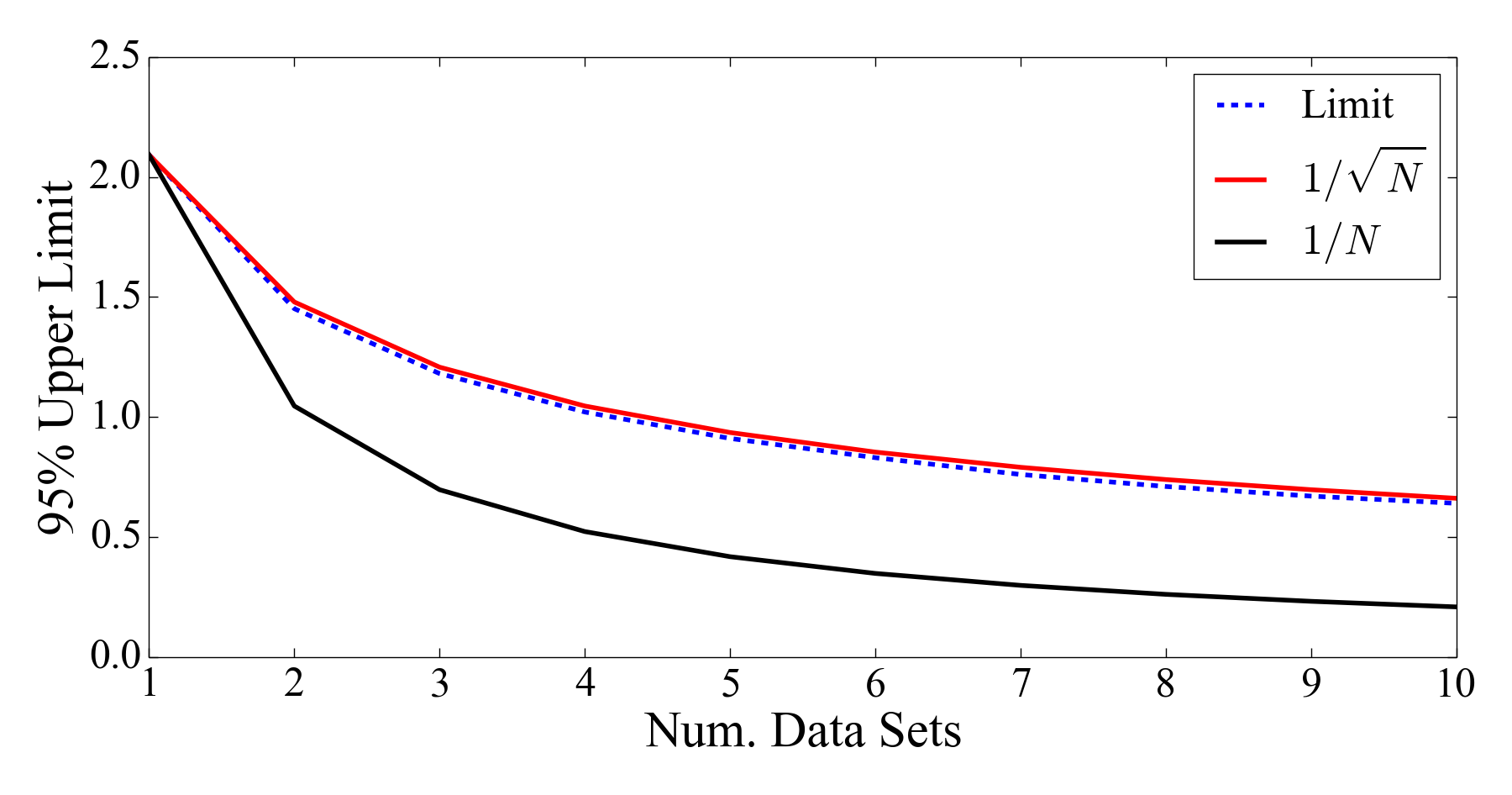}
\caption{Behavior of toy-model upper limits with the addition of sets.  Signal-to-background is 1:10 with 100 total events.\label{fig:run_targets}}
\end{figure}

Increasing the number of data sets comprising the joint likelihood naturally improves the power and tightens the limits, albeit at a rate dependent on their signal-to-noise ratios.  As long as the model uncertainties remain consistent, sets can be added indefinitely with no ill effect on the sensitivity.  As an example, see Fig. \ref{fig:run_targets}, where 95\% confidence upper limits are calculated with a cumulative number of toy-model sets.  Each set is identical, with signal-to-noise equal to 1:10 with 100 total events.  In this regime, limits improve with the square root of the number of sets, N.

In certain situations the joint upper limits can improve even more rapidly.  Any time $\rm ln[\mathcal{L}_{\rm joint}]\propto \mu$ holds throughout the allowed range of $\mu$, the constraints scale in direct proportion to N.  For example, a very low background might give a Poissonian likelihood, resulting in linear log-space behavior.  Forming the joint likelihood in log-space consists of adding these profiles together.  For the case of a set of linear functions, the limit level is then proportional to the sum of the slopes, i.e.

\begin{equation}
\mu_{\rm UL}\propto\left(\sum_{d}\frac{\partial \mathcal{L}_{d}}{\partial \mu}\Big|_{\mu=0}\right)^{-1}.
\end{equation}

\begin{figure}
\includegraphics[width=0.9\columnwidth]{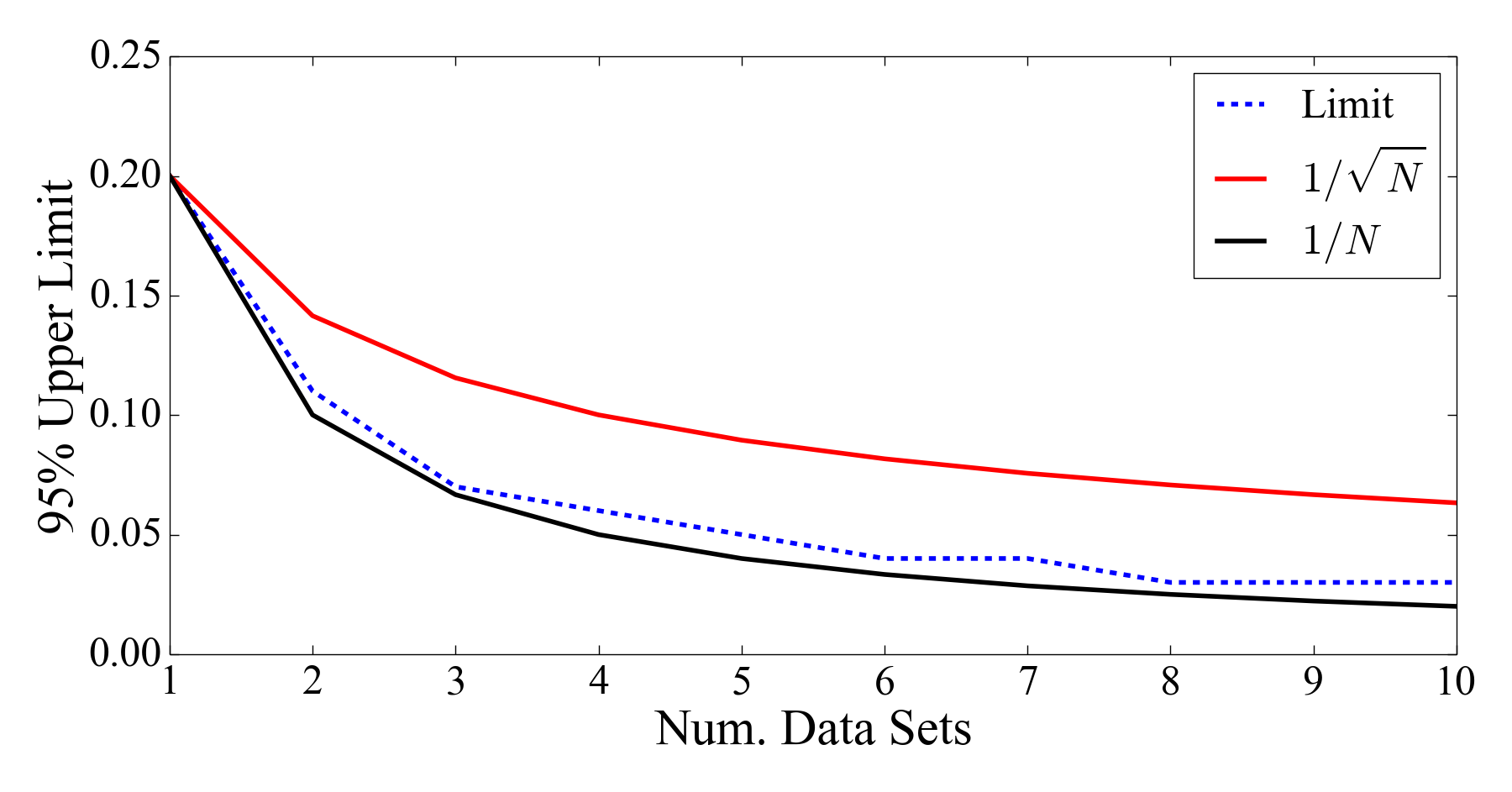}
\caption{Behavior of upper limits with the addition of data sets, with very low background [($s,b$) = (1,0.1)] and the constraint that $s>0$.\label{fig:run_targets_negMLE}}
\end{figure}

\noindent
The sum can be reduced to N in the case of set of profiles with identical slopes.  See Fig. \ref{fig:run_targets_negMLE} where a low background induces an appreciable effect.  

\section{Caveats}\label{sec:caveats}
\subsection{Overlapping Data Sets}
\begin{figure}
\includegraphics[width=0.9\columnwidth]{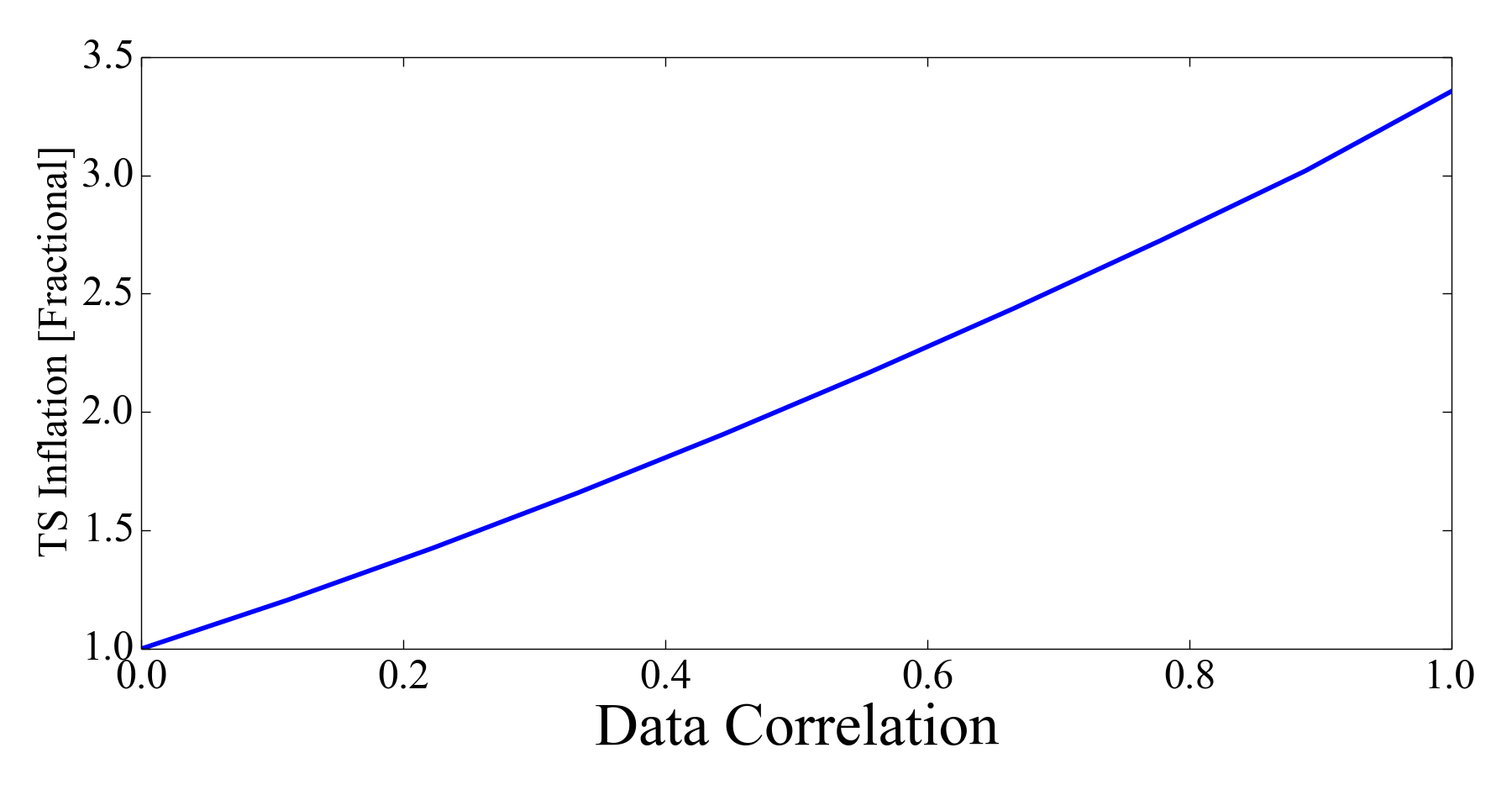}
\caption{The inflation of TS as two equivalent data sets gradually overlap.\label{fig:overlap}}
\end{figure}

\begin{figure}
\includegraphics[width=0.9\columnwidth]{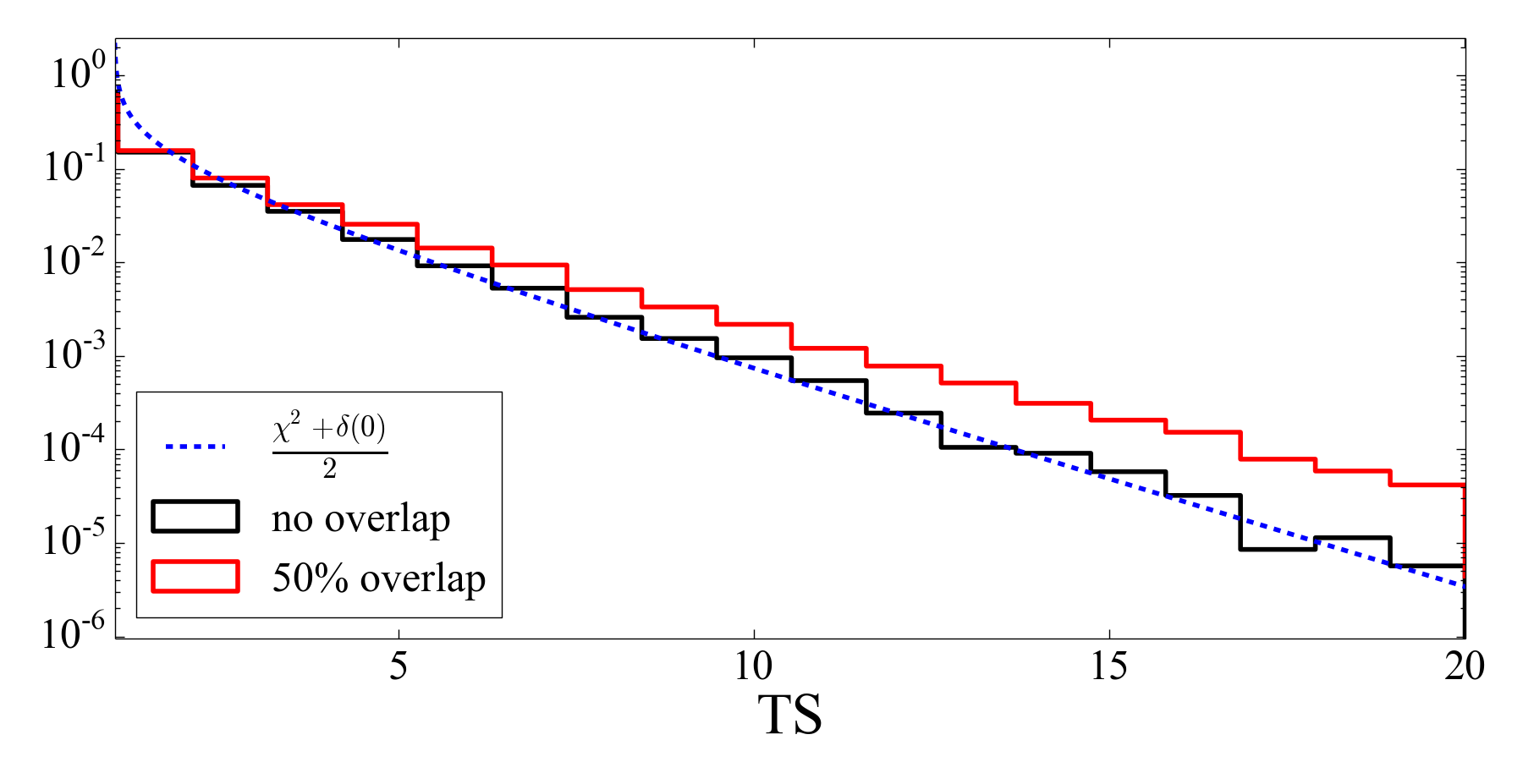}
\caption{Effect on the null distribution from a 50\% overlap correlation between a two-set joint likelihood.\label{fig:null_correlated}}
\end{figure}

It is best to avoid overlap between data sets.  If they do, then where there is signal, the TS will be erroneously increased by double-counting (Figure \ref{fig:overlap}), approximately in direct proportion to the percentage of overlap [See also appendix of \cite{2014ApJ...787...18A}].  When constructing a TS distribution, the significance derived from low-probability fluctuations will be similarly inflated.  See Figure \ref{fig:null_correlated}, where this is demonstrated using the preceding toy model.  
The upward skew there indicates that type II errors are more common than usual, effectively lowering the sensitivity of the study.  

 \section{Discussion and Conclusions}

The technique of joint likelihood, already widely used among Fermi-LAT Collaboration analyses, provides a straightforward and universal tool for combining constraints from astrophysical targets and other disjoint data sets.  We demonstrate that it matches the performance of residual stacking, and note that it often requires less effort to implement.  We model and describe the method's behavior in two interesting regimes:  first for very low background and second for the case of overlapping data sets.  The possible applications of the technique have by no means been exhausted and we encourage its continued use.  Lastly, we plan to expand on studies of the method's behavior in an upcoming publication.

\bigskip 
\begin{acknowledgments}
The authors wish to thank JACoW for their guidance in preparing
this template.

Work supported by Department of Energy contract DE-AC03-76SF00515.
\end{acknowledgments}

\bigskip 
\bibliography{bib}




\end{document}